\documentclass[twoside,twocolumn,9pt]{article}
\usepackage{extsizes}
\usepackage[super,sort&compress,comma]{natbib} 
\usepackage[version=3]{mhchem}
\usepackage[left=1.5cm, right=1.5cm, top=1.785cm, bottom=2.0cm]{geometry}
\usepackage{balance}
\usepackage{times,mathptmx}
\usepackage{sectsty}
\usepackage{graphicx} 
\usepackage{lastpage}
\usepackage[format=plain,justification=justified,singlelinecheck=false,font={stretch=1.125,small,sf},labelfont=bf,labelsep=space]{caption}
\usepackage{float}
\usepackage{fancyhdr}
\usepackage{fnpos}
\usepackage[english]{babel}
\usepackage{array}
\usepackage{droidsans}
\usepackage{charter}
\usepackage[T1]{fontenc}
\usepackage[usenames,dvipsnames]{xcolor}
\usepackage{setspace}
\usepackage[compact]{titlesec}
\usepackage{hyperref}

\usepackage{epstopdf}

\definecolor{cream}{RGB}{222,217,201}

\begin{document}

\pagestyle{fancy}
\thispagestyle{plain}
\fancypagestyle{plain}{

\fancyhead[C]{\includegraphics[width=18.5cm]{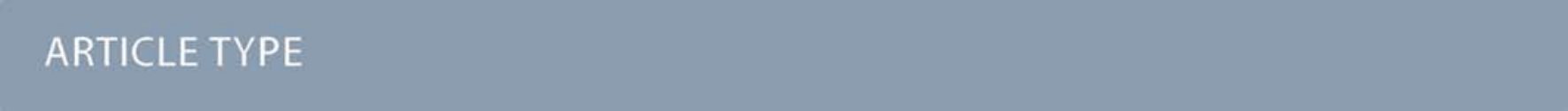}}
\fancyhead[L]{\hspace{0cm}\vspace{1.5cm}\includegraphics[height=30pt]{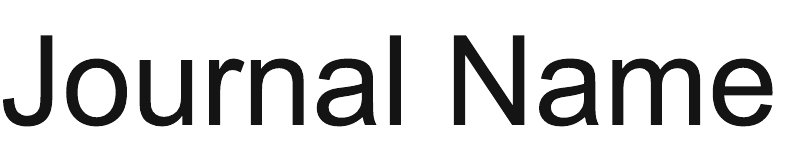}}
\fancyhead[R]{\hspace{0cm}\vspace{1.7cm}\includegraphics[height=55pt]{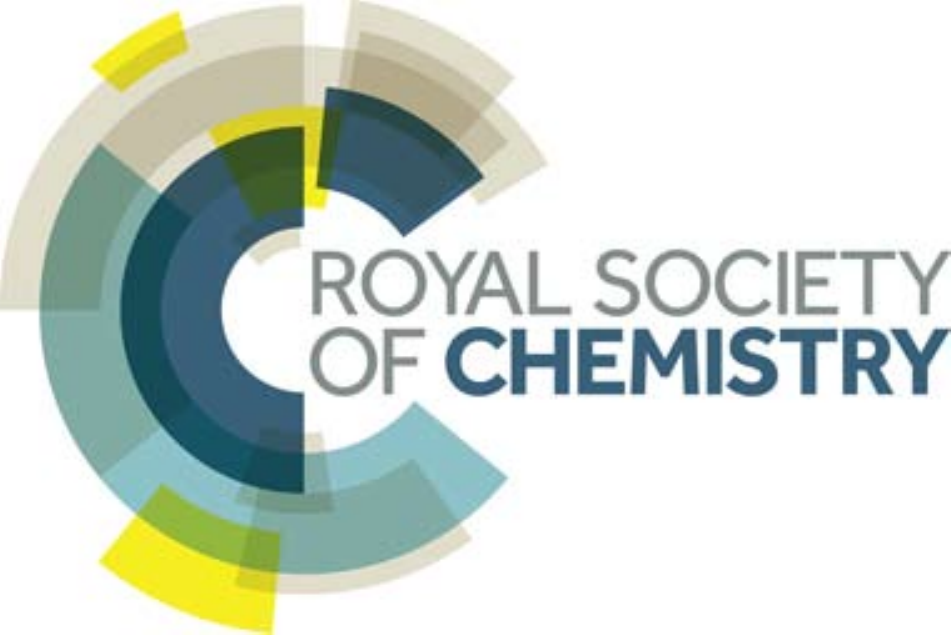}}
\renewcommand{\headrulewidth}{0pt}
}

\makeFNbottom
\makeatletter
\renewcommand\LARGE{\@setfontsize\LARGE{15pt}{17}}
\renewcommand\Large{\@setfontsize\Large{12pt}{14}}
\renewcommand\large{\@setfontsize\large{10pt}{12}}
\renewcommand\footnotesize{\@setfontsize\footnotesize{7pt}{10}}
\renewcommand\scriptsize{\@setfontsize\scriptsize{7pt}{7}}
\makeatother

\renewcommand{\thefootnote}{\fnsymbol{footnote}}
\renewcommand\footnoterule{\vspace*{1pt}%
\color{cream}\hrule width 3.5in height 0.4pt \color{black} \vspace*{5pt}} 
\setcounter{secnumdepth}{5}

\makeatletter 
\renewcommand\@biblabel[1]{#1}            
\renewcommand\@makefntext[1]%
{\noindent\makebox[0pt][r]{\@thefnmark\,}#1}
\makeatother 
\renewcommand{\figurename}{\small{Fig.}~}
\sectionfont{\sffamily\Large}
\subsectionfont{\normalsize}
\subsubsectionfont{\bf}
\setstretch{1.125} 
\setlength{\skip\footins}{0.8cm}
\setlength{\footnotesep}{0.25cm}
\setlength{\jot}{10pt}
\titlespacing*{\section}{0pt}{4pt}{4pt}
\titlespacing*{\subsection}{0pt}{15pt}{1pt}

\fancyfoot{}
\fancyfoot[LO,RE]{\vspace{-7.1pt}\includegraphics[height=9pt]{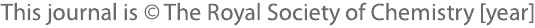}}
\fancyfoot[CO]{\vspace{-7.1pt}\hspace{13.2cm}\includegraphics{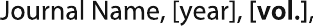}}
\fancyfoot[CE]{\vspace{-7.2pt}\hspace{-14.2cm}\includegraphics{head_foot/RF}}
\fancyfoot[RO]{\footnotesize{\sffamily{1--\pageref{LastPage} ~\textbar  \hspace{2pt}\thepage}}}
\fancyfoot[LE]{\footnotesize{\sffamily{\thepage~\textbar\hspace{3.45cm} 1--\pageref{LastPage}}}}
\fancyhead{}
\renewcommand{\headrulewidth}{0pt} 
\renewcommand{\footrulewidth}{0pt}
\setlength{\arrayrulewidth}{1pt}
\setlength{\columnsep}{6.5mm}
\setlength\bibsep{1pt}

\makeatletter 
\newlength{\figrulesep} 
\setlength{\figrulesep}{0.5\textfloatsep} 

\newcommand{\topfigrule}{\vspace*{-1pt}%
\noindent{\color{cream}\rule[-\figrulesep]{\columnwidth}{1.5pt}} }

\newcommand{\botfigrule}{\vspace*{-2pt}%
\noindent{\color{cream}\rule[\figrulesep]{\columnwidth}{1.5pt}} }

\newcommand{\dblfigrule}{\vspace*{-1pt}%
\noindent{\color{cream}\rule[-\figrulesep]{\textwidth}{1.5pt}} }

\makeatother

\twocolumn[
  \begin{@twocolumnfalse}
\vspace{3cm}
\sffamily
\begin{tabular}{m{4.5cm} p{13.5cm} }

\includegraphics{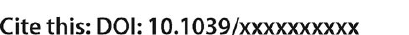} & \noindent\LARGE{\textbf{Compositional nanodomain formation in hybrid formate perovskites$^\dag$}} \\
 & \vspace{0.3cm} \\

 & \noindent\large{Edwina A. Donlan, Hanna L. B. Bostr{\"o}m, Harry S. Geddes, Emily M. Reynolds, and Andrew L. Goodwin$^{\ast}$} \\

\includegraphics{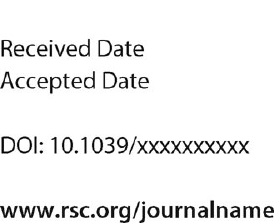} & \\

\end{tabular}

 \end{@twocolumnfalse} \vspace{0.6cm}

  ]

\renewcommand*\rmdefault{bch}\normalfont\upshape
\rmfamily
\section*{}
\vspace{-1cm}


\footnotetext{\textit{Department of Chemistry, University of Oxford, Inorganic Chemistry Laboratory, South Parks Road, Oxford OX1 3QR, U.K. Tel: +44 (0)1865 272137; E-mail: andrew.goodwin@chem.ox.ac.uk}}

\footnotetext{\dag~Electronic Supplementary Information (ESI) available: Experimental methods; powder diffraction refinement details; lattice parameters; semaphore diagram descriptions; IR spectra; NMF fits; equilibrium calculations and RMC refinement results. See DOI: 10.1039/b000000x/}




\sffamily{\textbf{We report the synthesis and structural characterisation of three mixed-metal formate perovskite families [C(NH$_{\textbf 2}$)$_{\textbf 3}$]M$_{\textbf {1}\boldsymbol{-x}}$Cu$_{\boldsymbol x}$(HCOO)$_{\textbf 3}$ (M = Mn, Zn, Mg). Using a combination of infrared spectroscopy, non-negative matrix factorization, and reverse Monte Carlo refinement, we show that the Mn- and Zn-containing compounds support compositional nanodomains resembling the polar nanoregions of conventional relaxor ferroelectrics. The M = Mg family exhibits a miscibility gap that we suggest reflects the limiting behaviour of nanodomain formation.}}\\


\rmfamily 


Compositional heterogeneity is an essential ingredient in a number of important classes of functional materials.\cite{ref1} Arguably the clearest example is that of the relaxor ferroelectrics, such as $(1-x)$Pb(Mg$_{1/3}$Nb$_{2/3}$)O$_3$--$x$PbTiO$_3$ (PMN-PT):\cite{ref2} here an inhomogeneous distribution of B-site cations allows the formation of polar nanoregions (PNRs),\cite{ref3,ref4,ref5} the collective motion of which is responsible for the giant electromechanical response observed and exploited experimentally.\cite{ref6,ref7} Recently, the same ideas have been extended to relaxor ferromagnets,\cite{ref8} which are in turn conceptually related to cluster spin glasses,\cite{ref9} known for their exotic magnetic memory effects.\cite{ref10} Likewise, in some porous metal--organic frameworks (MOFs), inhomogeneous vacancy distributions\cite{ref11} affect gas uptake\cite{ref12} via a mechanism that is analogous to ion conduction pathways in compositionally-heterogeneous solid-oxide fuel-cell materials.\cite{ref13}

From a materials design perspective, the existence of nanoscale compositional inhomogeneities relies on a delicate balance of interaction strengths at the atomic scale. If the interaction between different components is essentially independent of composition then mixing will be uniform; this is nearly always the case for rare-earth substitution in garnets, for example.\cite{ref14} Conversely, if interaction strengths vary too greatly then either macroscopic phase segregation (like interactions favoured, \emph{e.g.}\ eutectics\cite{ref15}) or component ordering (unlike interactions favoured, \emph{e.g.}\ double perovskites\cite{ref16,ref17}) occurs. So the task of establishing chemical control over nanodomain structures is an important challenge in the field.\cite{ref18,ref19,ref20}

\begin{figure}[t]
\centering
  \includegraphics{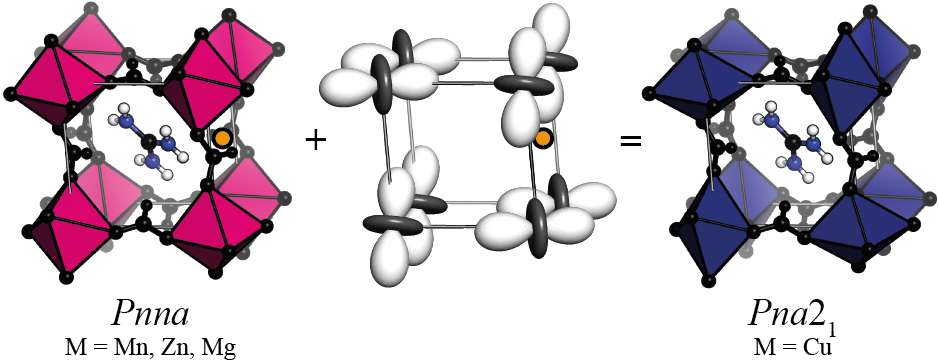}
  \caption{The crystal structures of most [Gua]M(HCOO)$_3$ perovskite analogues (left) are non-polar. Collective JT order (centre) breaks inversion symmetry\cite{ref21,ref22} (\emph{e.g.}\ at the site marked by an orange circle). Hence [Gua]Cu(HCOO)$_3$ (right) is polar, with the orientation of polarization a function of the phase of collective JT order.\cite{ref21}}
  \label{fig1}
\end{figure}

It was in this context that we sought to explore cation distributions in the mixed-metal formate perovskites [Gua]Cu$_x$M$_{1-x}$(HCOO)$_3$ (Gua$^+$ = C(NH$_2$)$_3^+$; M = Mn, Zn, Mg). It has been known for some time that many formate perovskites show relaxor-like dielectric behaviour,\cite{ref23,ref24,ref25} which is usually associated with glassy dynamics of the organic (A-site) cation.\cite{ref26,ref27} Indeed a recent $^{13}$C NMR study of [(CH$_3$)$_2$NH$_2$]Zn(HCOO)$_3$ even demonstrated the spontaneous formation of fluxional PNRs during the onset of ferroelectric order.\cite{ref28} By contrast, the exploration of mixed-metal formate perovskites is rather less well developed.\cite{ref29,ref30,ref31,ref32,ref33} Our focus here on the [Gua]M(HCOO)$_3$ family is motivated by the link in this particular system between polarization and cooperative Jahn--Teller (JT) order for M = Cu [Fig.~\ref{fig1}].\cite{ref21,ref22,ref30} In particular, we anticipated that substitution on the Cu site by JT-inactive cations (\emph{i.e.}, Mn, Zn, Mg; note Cd gives a different structure type\cite{ref30,ref34}) might be expected to reduce the length-scale of polar order, and---if clustering were found to occur---then favour the formation of PNRs in Cu dilute compositions. Moreover, a computational study of [Gua]Mn$_{0.5}$Cu$_{0.5}$(HCOO)$_3$ has recently suggested strong enhancement in both polarization and magnetization relative to the Mn- and Cu-containing end-members.\cite{ref35}

Our focus here is on the synthesis and structural characterisation of three [Gua]Cu$_x$M$_{1-x}$(HCOO)$_3$ families (M = Mn, Zn, Mg) rather than their dielectric behaviour, (although we anticipate our results will have strong implications for the latter). Making use of a newly-developed non-negative matrix factorization\cite{ref36,ref37} (NMF) analysis of infrared (IR) spectroscopy data, we determine the experimental distribution of neighbouring cation pairs as a function of composition. We use these data to drive reverse Monte Carlo\cite{ref38} (RMC) refinements of cation distributions, which in turn reveal nanodomain formation precisely of the type observed in conventional relaxor ferroelectrics such as PMN-PT.

\begin{figure}[t]
\centering
  \includegraphics{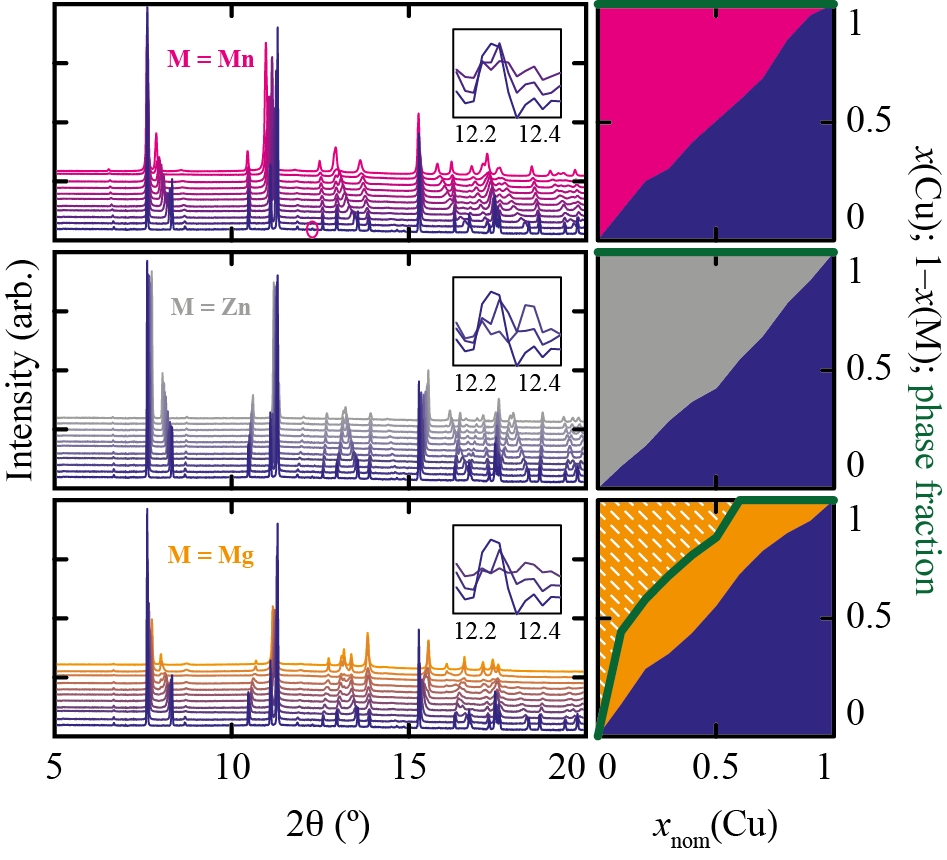}
  \caption{Variation in synchrotron X-ray powder diffraction patterns (left, $\lambda$ = 0.82598\,\AA) and phase fractions/compositions (right) for [Gua]Cu$_x$M$_{1-x}$(HCOO)$_3$ with nominal Cu content $x_{\textrm{nom}}$(Cu) = 0--1 (top--bottom in left panels). Insets show the strongest $Pna2_1$ superlattice reflection (circled) for $x$ = 0.8, 0.9, 1. For M = Mn, Zn, $x$(Cu) values were determined by AAS. For M = Mg, compositions (represented by division of block colour) and phase fractions (green lines) were determined by Rietveld refinement. For $x <$ 0.6, the dashed component represents [Gua]Mg(HCOO)$_3$ and the solid region [Gua]Cu$_x$Mg$_{1-x}$(HCOO)$_3$ (see SI for analysis and further discussion).}
  \label{fig2}
\end{figure}

Polycrystalline samples of [Gua]Cu$_x$M$_{1-x}$(HCOO)$_3$ (M = Mn, Zn, Mg; $x$ = 0, 0.1, $\ldots$ , 1) were prepared according to the method of Ref.~\citenum{ref39}; to the best of our knowledge neither the mixed-cation nor the Mg-endmember frameworks have been reported.\cite{ref34,ref39} Cu:M ratios in our samples were determined by atomic absorption spectroscopy (AAS), and synchrotron X-ray powder diffraction measurements used to establish the phase behaviour of each family (see SI for details). We find that the M = Mn and Zn families form continuous solid solutions, whereas the M = Mg system precipitates as a two-phase mixture for $x <$ 0.6 [Fig.~\ref{fig2}]. Rietveld refinement of these diffraction data indicates that Cu has negligible solubility in [Gua]Mg(HCOO)$_3$, but that Mg has some limited solubility ($<$ 50\%) in [Gua]Cu(HCOO)$_3$.

Weak superlattice reflections associated with collective JT order are evident in the diffraction patterns of the phases with high Cu contents [Fig.~\ref{fig2}]. In principle, this part of the diffraction pattern contains information regarding the extent and nature of polar domain formation: either via Bragg intensities in the case of long-range JT order, or \emph{via} diffuse scattering if order is short-range.\cite{ref40,ref41} The difficulty in this case is that the superstructure scattering intensity is so weak that we could not draw any unambiguous conclusions regarding cation distributions on the basis of these diffraction patterns.

\begin{figure}[t]
\centering
  \includegraphics{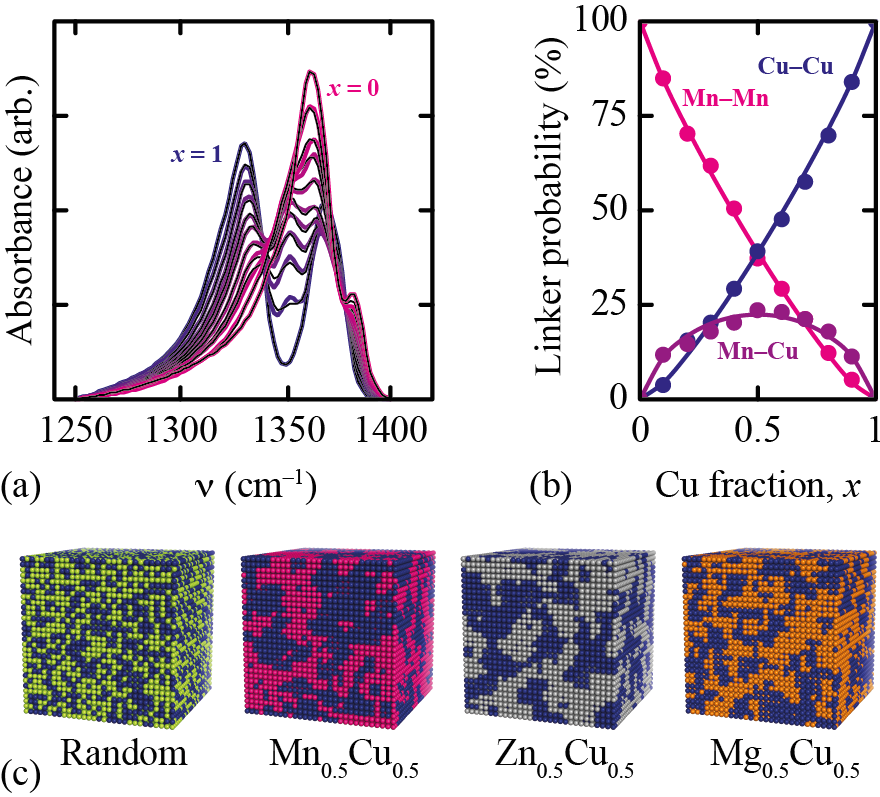}
  \caption{(a) Normalised IR absorption spectra of [Gua]Cu$_x$Mn$_{1-x}$(HCOO)$_3$ in the formate C--O stretching region (coloured lines) with NMF fits in black. (b) NMF linker distributions (data points) with equilibrium fit ($\Delta H_{\textrm{mix}}$ = 3.1\,kJ\,mol$^{-1}$) as a solid line. (c) Representative $x$ = 0.5 RMC configurations for M = Mn, Zn, and Mg, showing nanoscale clustering of Cu-rich regions (dark blue).}
  \label{fig3}
\end{figure}

Consequently, we turned to IR spectroscopy, focusing on the 1250--1400\,cm$^{-1}$ region because it contains the carboxylate stretches of the bridging formate.\cite{ref39} We rationalised that formate stretching frequencies should be sensitive to the nature of the two cations to which a formate anion is bound, and hence allow characterisation of the distribution of Cu--Cu, Cu--M, and M--M neighbours in our samples. These distributions, in turn, reflect the degree of M/Cu mixing and hence the presence or absence of any nanodomain formation. Our results show a continuous variation in IR absorption profile as a function of Cu content $x$ [Fig.~\ref{fig3}(a)]. At intermediate values of $x$, features appear that are not observed in either endmember; we attribute these to stretches of formate anions that bridge Cu--M pairs. 

We used a custom implementation of the NMF algorithm\cite{ref36,ref37} (see SI) to deconvolve each IR absorption profile into three components: two corresponding to the profiles for each endmember (\emph{i.e.}\ M--M and Cu--Cu) and the third corresponding to that of the mixed-metal linkages (M--Cu). In this way, the relative fraction of M--M, M--Cu, and Cu--Cu linkages can be obtained across each series. In Fig.~\ref{fig3}(b) we show the results of our analysis for M = Mn; those for M = Zn and Mg, which are entirely similar, are given in the SI.

Our key result is that the distribution of these different linker types indicates cation segregation at the atomic scale. Taking the $x$ = 0.5, M = Mn composition as an example, a statistical distribution of Mn and Cu cations would give Mn--Mn:Mn--Cu:Cu--Cu ratios of 25:50:25; we find 39:24:37. Indeed, for all compositions, and for all choices of M, we observe much smaller mixed-cation M--Cu fractions than would be expected for a random distribution. Assuming equilibrium conditions, and taking into account the entropy of mixing contribution to the free energy, we can use these linker distributions to estimate the enthalpy penalty of cation mixing: we find $\Delta H_{\textrm{mix}}$ = 3.1 and 3.6\,kJ per mole of cation--cation interactions for M = Mn and Zn (the corresponding value for Mg, 2.4\,kJ\,mol$^{-1}$, is much less reliable as a result of the presence of two phases). By way of context, we note that these values are significantly larger than those of mixed-cation hybrid lead halide perovskites.\cite{ref42} 

In order to understand the microscopic implications of the linker distributions determined in our IR/NMF analysis, we used an RMC approach.\cite{ref38} The basic idea was to generate an atomistic supercell containing a fixed number of M and Cu sites according to the composition of interest. The fraction of M--M, M--Cu, and Cu--Cu neighbours can be calculated straightforwardly from such a model. We then used the Metropolis Monte Carlo algorithm\cite{ref43} to swap M and Cu atoms until the configuration reproduced the experimental linker distributions. Hence the resulting RMC configurations are representative states consistent with experimental IR data, and are physically meaningful irrespective of the precise extent to which the experimental systems are at equilibrium. Configurations for all compositions are given as SI, but we illustrate the specific case of $x$ = 0.5 in Fig.~\ref{fig3}(c). What is immediately clear is that considerable nanoscale segregation is evident in all three families, with the degree of segregation reflecting the magnitude of $\Delta H_{\textrm{mix}}$. Indeed it is possible to identify nanodomains of the type associated with relaxor PNRs, a result we have obtained without assumptions regarding nanodomain shape.\cite{ref44}

That the M = Mg system forms a two-phase mixture rather than a solid solution indicates the delicate energy balance required to support nanodomain formation in these systems. We suggest that the particular propensity for phase separation in [Gua]Cu$_x$Mg$_{1-x}$(HCOO)$_3$ reflects the large difference in end-member unit-cell metric for this system. The relative ordering of $\Delta H_{\textrm{mix}}$ for M = Mn and Zn can also be rationalised on the basis of strain arguments (see SI). Variation in synthesis temperature (all our samples were prepared under ambient conditions) may offer a route to stabilise a continuous solid solution for the M = Mg phase, and indeed allow further control over nanodomain size in all three families.

To the best of our knowledge, the only other MOF family in which compositional nanodomains have been characterised experimentally is that of UiO-66.\cite{ref11} In that particular case, the key implications of nanodomain formation are porosity (hence adsorption profile) and mechanical response, including thermal expansion.\cite{ref12,ref45} For the formate perovskites we study here, the observation of Cu clustering has important implications for the formation of PNRs and hence relaxor behaviour. The emergence of polarisation on the nanometer scale requires local JT order within Cu-rich domains that is of the same type as that observed in [Gua]Cu(HCOO)$_3$ itself. Computational studies of other dilute JT systems suggest this to be likely;\cite{ref46} moreover, even in the presence of a percolating network of Cu sites one expects decoupling of the phase of JT order (and hence polarisation direction) between connected Cu-rich regions as a result of additional JT degrees of freedom in Cu-poor regions.\cite{ref47} This decoupling---which is key for relaxor behaviour---is consistent with our experimental observation that long-range JT order [Fig.~\ref{fig2}] disappears much more quickly on doping than expected from percolation considerations alone.\cite{ref30,ref47} So our study has clearly identified these families of mixed-cation formate perovskites as obvious relaxor candidates. Experimental studies of the dielectric response of these materials are a natural avenue of future research; we suggest that the Cu-rich compositions without long-range JT order are likely to exhibit the most interesting behaviour.

\section*{Acknowledgements}
The synchrotron diffraction measurements were carried out at the Diamond Light Source (I11 Beamline). We are extremely grateful for the award of a Block Allocation Grant, which made this work possible, and for the assistance in data collection provided by M. S. Senn (Warwick), C. S. Coates (Oxford) and the I11 beamline staff. We acknowledge studentship from the E.P.S.R.C. and the University of Oxford to H.S.G. and H. L. B., respectively. This project received funding from the European Union (EU) Horizon 2020 Research and Innovation Programme under Marie Sklodowska-Curie Grant Agreement 641887 (project acronym DEFNET).

\section*{Conflict of interest}
There are no conflicts to declare.



\balance


\scriptsize{
\bibliography{cc_2017_nanodomain} 
\bibliographystyle{rsc} } 

\end{document}